\documentstyle[epsf,rotate]{elsart}

\begin{document}

\begin{frontmatter}

\title{Directed percolation with an absorbing boundary}

\author[NBI]{Kent B{\ae}kgaard Lauritsen\thanksref{KBL}}
\author[NORDITA]{Kim Sneppen}
\author[NBI,SK]{ Maria Marko\v sov\'a}
\author[NBI]{ Mogens H. Jensen}

\thanks[KBL]{baekgard@nbi.dk}

\address[NBI]{Niels Bohr Institute, Center for Chaos and Turbulence Studies, 
        Blegdamsvej 17, 2100 Copenhagen \O, Denmark}
\address[NORDITA]{NORDITA, Blegdamsvej 17, 2100 Copenhagen \O, Denmark}
\address[SK]{Institute of Measurement Science, Slovak Academy of Science, 
	 Dubravska cesta 9, 842 19 Bratislava, Slovakia}

%
%
%
%

\date{\today}

\begin{abstract}
We consider directed percolation with an absorbing boundary
in $1+1$ and $2+1$ dimensions.
The distribution of cluster lifetimes and sizes depend on the boundary.
The new scaling exponents can be related to the exponents
characterizing standard directed percolation in $1+1$ dimension.
In addition, we investigate the backbone cluster and red bonds, and
calculate the distribution of living sites along the
absorbing boundary.
\end{abstract}


\end{frontmatter}


%
%
%


\section{Introduction}

Directed percolation (DP) is the common name for 
spreading processes with active (``live'') sites and
an absorbing (``dead'') state.
The process of directed percolation is more precisely defined on a lattice
where at each time step a site can become alive with probability
$p$, if and only if at least one of 
its neighbors was alive at the previous time step \cite{schloegl}.
For $p\rightarrow 0$ the process rapidly terminates whereas
for $p\rightarrow 1$ the live sites spread without limit.
There exists a critical value $p=p_c$ such that the
distribution of clusters originating from an initial live site is a power law.
The probability that there are living sites
at time $t$ after the introduction of one living site
in a infinite lattice of dead sites at time $t=0$ is
\cite{annphys}
\begin{equation}
	P(t) \sim t^{-\delta}, \qquad \delta = \beta/\nu_{\parallel} .
					\label{eq:P(t)}
\end{equation}
Here $\beta$ is the order parameter exponent
which determines the probability $\epsilon^\beta$
($\epsilon \equiv p-p_c$) for the cluster to survive to infinity,
and $\nu_{\parallel}$ and $\nu_{\perp}$ are
the correlation length exponents along
($\xi_\parallel \sim \epsilon^{-\nu_\parallel}$)
and perpendicular
($\xi_\perp \sim \epsilon^{-\nu_\perp}$)
to the time direction of the DP process.
Both $\beta$ and $\nu_{\parallel}$ enter in Eq.~(\ref{eq:P(t)})
which reflect the fact that the final survival of the cluster is
determined when the cluster size reaches the correlation length 
$\xi_{\parallel}$; the probability to reach $\xi_{\parallel}$
scales as $\epsilon^{\beta} \sim \xi_{\parallel}^{-\beta/\nu_{\parallel}}$
in accordance with Eq.~(\ref{eq:P(t)}).

Directed percolation is known to be equivalent to
Reggeon field theory \cite{rft,rft2} which describes the evolution of
a density $\rho$ of live sites for a large class of processes.
An understanding of facets of directed percolation 
have implications for the understanding of a number of
widely different problems in physics,
including the dynamics of chemical reactions and catalyzers
\cite{schloegl,annphys}, contact processes \cite{liggert},
spatio-temporal intermittency \cite{pomeau,chate,rolf-etal},
self-organized criticality \cite{soc,soc2,huber,maria},
and directed polymers \cite{zhang}.

In the present paper we consider directed percolation
with an absorbing boundary \cite{janssen-etal,essam-etal:1996}.
In the context of spatio-temporal intermittency
\cite{pomeau,chate,rolf-etal},
the process of directed perclation can be viewed as lateral convection
of intermittent spots along a boundary which is absorbing
(i.e., the boundary layer has laminar flow). Thus the active
sites represent intermittent spots that originate from one
point on the boundary due to a impurity or an edge. The deacy
of the turbulent spots into the absorbing (laminar flow) could
possibly be connected to directed percolation with a wall.
Another more traditional picture is a catalytic process
that is initiated at the edge of the catalyzer. 
As a result, the propagation of active live sites can then
only propagate into the system. We consider very slow
activation on the boundary: the activity in the system is
supposed to be entirely determined by the propagation
of one initial spark at the boundary at time zero.

\begin{figure}[htb]
\centerline{
        \epsfxsize=9.0cm
        \epsfbox{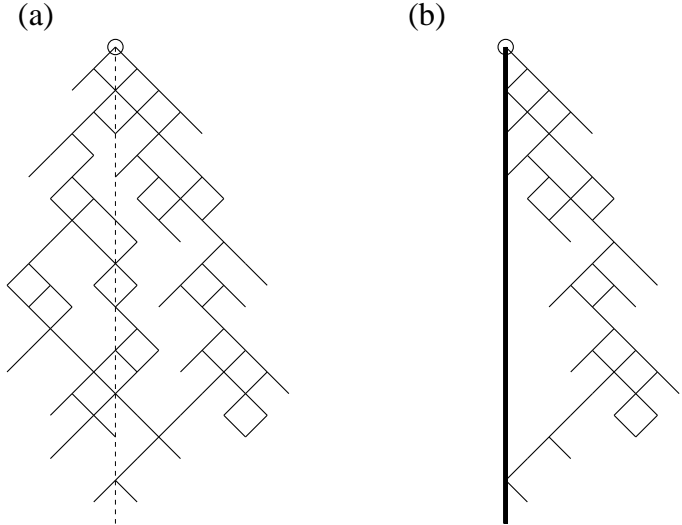}
}
\vspace*{0.5cm}
\caption{a) 
	Schematic drawing of a DP cluster.
	b) 
	Schematic drawing of the same DP cluster but now with the
	absorbing boundary present. One notices the existence of 
	a different distribution of voids compared to the case (a).
	The time axis is directed downwards.
	Note that in our implementation of the absorbing boundary we allow
	for activity on the boundary; not allowing for this
	simply amounts to shifting the absorbing boundary one
	lattice spacing to the left.}
\vspace*{0.5cm}
\label{fig:clusters}
\end{figure}

We consider time-directed percolation in a geometry as 
shown in figure~\ref{fig:clusters}, and will refer to the $d$ transverse
direction as space ($x$) and to the longitudinal direction
as time ($t$).
The boundary absorbs the parts of the DP cluster which
in an open geometry would have continued spreading on the other side.
We will first discuss the $1+1$-dimensional ($d=1$) directed site DP
on a square lattice where $p_c=0.705485$, $\beta=0.2765$,
$\nu_\parallel=1.7338$, and $\nu_\perp=1.0968$
\cite{essam-etal:1996,1d,1d-2,iwan};
in the following,
the subscript '1' denotes critical exponents for DP with an
absorbing boundary.
Recently, series expansions for $1+1$-dimensional
bond directed percolation in a similar
absorbing geometry were carried out \cite{essam-etal:1996}.
Our results in $d=1$ are in complete agreement with the estimates in
\cite{essam-etal:1996}.

\section{Critical behavior}

\begin{figure}[tb]
\centerline{
\epsfysize=.8\columnwidth{\rotate[r]{\epsfbox{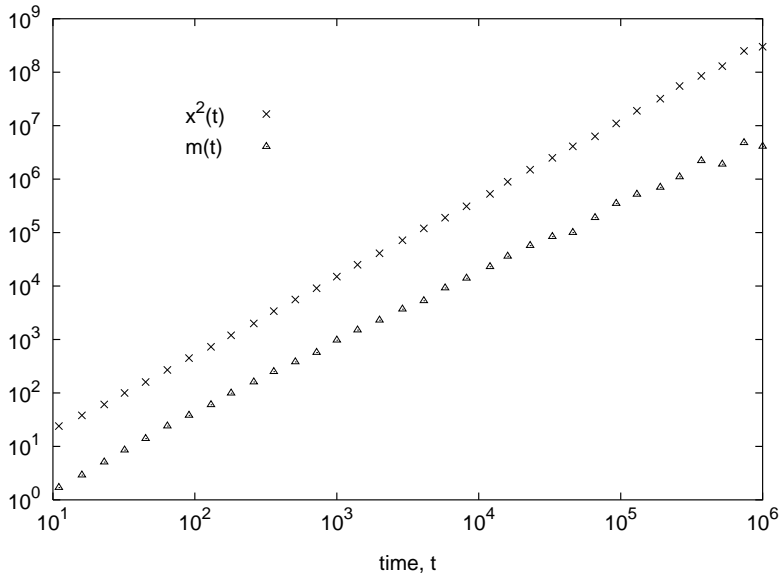}}}
}
\vspace*{0.5cm}
\caption{Cluster mass, $m(t)$, and meandering, $x^2(t)$, as function of time
	in $1+1$ dimensions.
	We obtain the mass scaling exponent 
	$1+(\nu_{\perp}-\beta)/\nu_{\parallel} = 1.475 \pm 0.005$,
	and the meandering exponent 
	$2\chi \equiv 2\nu_{\perp}/\nu_{\parallel} = 1.265 \pm 0.005$,
	in agreement with the values for DP without an absorbing boundary.
	}
\vspace*{0.5cm}
\label{fig:mass}
\end{figure}

First we notice that there remain a critical value
$p_c$ for DP with an absorbing boundary,
and that this value is independent of the presence of such a boundary. 
In figure~\ref{fig:mass} we display the
meandering (squared) $x^2(t)$ and the mass $m(t)$ of the cluster
measured for a value of $p$ close to $p_c$
($\epsilon = -0.0015$).
The meandering scales as $\xi_{\perp}^{2} \sim t^{2\nu_\perp/\nu_\parallel}$,
with $t$ determined by $\xi_\parallel$.
The mass scales as $\xi_\parallel \xi_\perp \epsilon^\beta
\sim t^{1+(\nu_{\perp}-\beta)/\nu_{\parallel}}$.
We find that both these quantities scale in the same way as if there
was no boundary. Thus, both the meandering exponent 
$\chi \equiv \nu_{\perp}/\nu_{\parallel}=0.6327$ 
and the mass scaling exponent 
$1+(\nu_{\perp}-\beta)/\nu_{\parallel} = 1.4732$ are 
independent of the absorbing boundary.
We also measure the lifetime distribution and average lifetime
(see below) and conclude that the three exponents
($\beta$, $\nu_{\parallel}$, and $\nu_\perp$)
for DP are still present for DP with the presence of a wall.
This is consistent with the fact that DP clusters with
a wall consist of subsets of DP clusters without a wall.
Therefore the correlation length and thus the
correlation length exponents should be 
identical in the two cases. 
Further, for the subset of clusters that 
survive to a given time $t$, there exist a connected
leftmost path closest to the wall that merge the starting point with a point
that survive to time $t$. To the right of this path the cluster is identical
to a normal DP cluster with the same scaling behavior of the density.
The fact that the scaling of the
mass of a DP cluster with and without a wall are identical show 
that for the subset of clusters that survive to a given time, the 
active points within these clusters are on average so far from
the wall that the density is not influenced by the presence of the wall.

\begin{figure}[htb]
\centerline{
\epsfysize=.8\columnwidth{\rotate[r]{\epsfbox{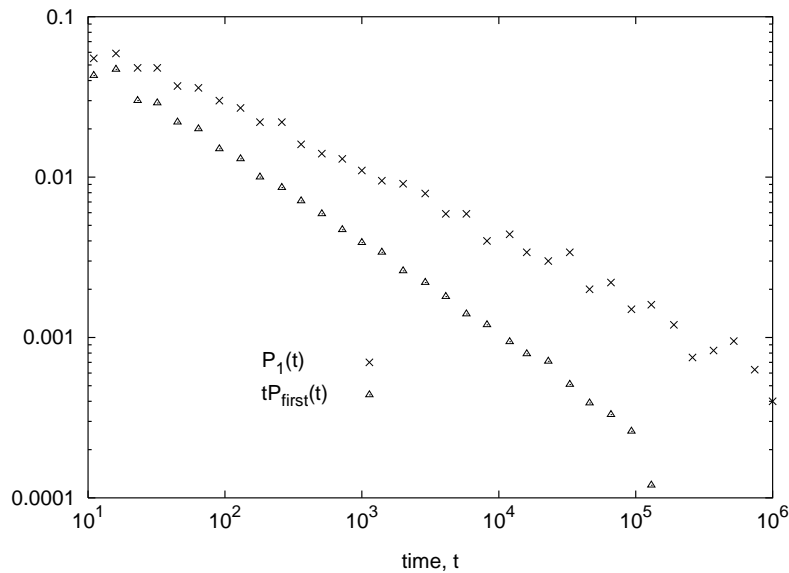}}}
}
\vspace*{0.5cm}
\caption{Probability $P_1(t)$ that there are still live sites at time $t$
	in $1+1$ dimensions. We obtain
	a power law with exponent $\delta_1 = 0.42 \pm 0.01$.
	We also display the distribution $p_{\rm first}(t)$ of first return 
	along the absorbing boundary and obtain a power-law distributions 
	with exponent $\tau_{\rm{first}}=1.577 \pm 0.005$.
	}
\label{fig:boundary}
\end{figure}

\begin{figure}[htb]
\centerline{
\epsfysize=.8\columnwidth{\rotate[r]{\epsfbox{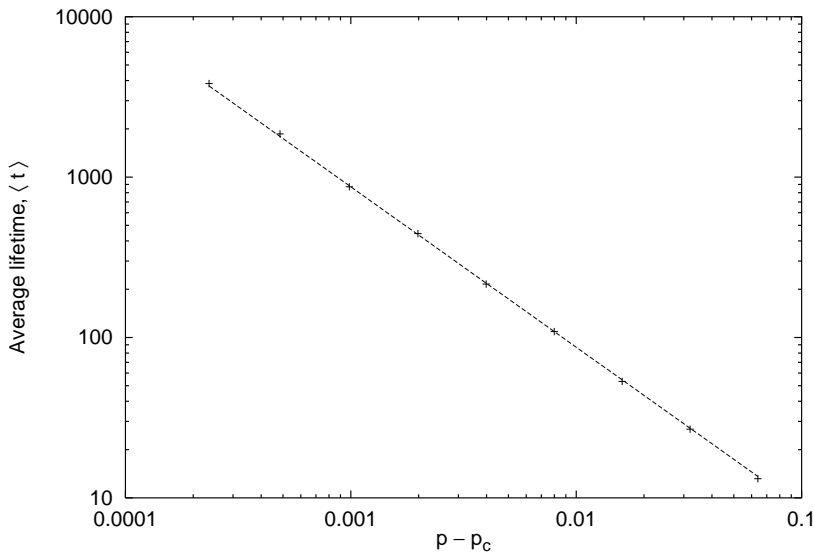}}}
}
\vspace*{0.5cm}
\caption{Average lifetime as function of $p$ for values above $p_c$
	in $1+1$ dimensions.
	The estimate for the slope is $\tau_1 = 1.000 \pm 0.005$.
	}
\vspace*{0.5cm}
\label{fig:<t>}
\end{figure}

Next, we investigate the distribution of lifetimes
in the case of an absorbing boundary. In figure~\ref{fig:boundary}
we show the probability that there are still living sites at time $t$.
We find that the presence of an absorbing boundary 
changes the exponent of the lifetime distribution from
the one given by~(\ref{eq:P(t)}) (i.e., $\delta=0.1594$) to 
\begin{equation}
	P_1(t) \sim t^{-\delta_1}, \qquad \delta_1 = 0.420 \pm 0.005.  
					\label{eq:P_1(t)}
\end{equation}
We have $\delta_1=\beta_1/\nu_\parallel$,
where the order parameter exponent $\beta_1$ determines
the probability $\epsilon^{\beta_1}$ for the cluster to survive to
infinity. This yields the value $\beta_1 = 0.728 \pm 0.008$,
and we conclude that $\beta_1 \neq \beta$. Note, as mentioned, that $\beta$
is still needed for DP with a boundary
in order to describe the scaling of the cluster mass
near $p=p_c$.  {}From the full scaling behavior 
\begin{equation}
	P_1(t) = t^{-\delta_1} f(t \epsilon^{\nu_\parallel}) , 
					\label{eq:P_1(t)-full}
\end{equation}
we obtain the average lifetime 
\begin{equation}
	\left< t \right> \sim \epsilon^{-\tau_1}, 
			\qquad 
			\tau_1 = \nu_\parallel (1-\delta_1)
			         = \nu_\parallel -\beta_1  .
						\label{eq:<t>}
\end{equation}
In Ref.\ \cite{essam-etal:1996}
they obtain $\tau_1=1.000...$ and conjecture that $\tau_1$
is exactly equal to unity. This leads to 
$\beta_1 = \nu_\parallel - 1$, and
$\delta_1 = 1-1/\nu_\parallel = 0.4232$ in agreement with our result.
We also measure $\left< t \right>$ directly (Fig.~\ref{fig:<t>}) and
obtain the estimate $\tau_1 = 1.000 \pm 0.005$,
yielding an estimate for $\beta_1$ in agreement with the one above.
Even though we do not know of an analytical argument for $\tau_1=1$,
we can obtain an upper bound by the following heuristic
argument: if we assume that the statistics of the DP cluster are
described by the statistics of the ``outer envelope'' which scales
as $t^\chi$ we obtain that the first return of the envelope back
to the absorbing boundary determines the cluster lifetime distribution.
The probability that the envelope returns to ``zero''
(i.e., to the absorbing boundary) is accordingly given by the scaling
of the first return of a fractional Brownian motion with exponent $\chi$:
\begin{equation}
	p'(t) = \frac{1}{t^{2-\chi}}  .
					\label{eq:p_chi}
\end{equation}
This yields the survival exponent
$\delta' = (2 - \chi) - 1 = 1 - \nu_\perp/\nu_\parallel$,
and thus the estimate $\tau_{1}' = \nu_\perp$,
only a little larger than the correct value.
The reason that the `envelope' argument is only approximate is due to
the fact that clusters can spontaneously die out before the envelope
returns to the boundary.

Our results for the probability $P_1(s)$ to have a cluster of size
larger than $s$ show that it follows
a power law with an exponent $\tau_s -1 = 0.283 \pm 0.005$.
This value can easily be explained in terms of the lifetime
distribution investigated above. A cluster of lifetime $t$ will
have a size (mass) $s$ of the order
$s \sim t^{1+(d\nu_\perp-\beta)/\nu_\parallel}$,
where the density exponent $\beta$ enters (and not the order parameter
exponent $\beta_1$ associated with a seed on the boundary).
Transforming the probability distribution (\ref{eq:P_1(t)-full}) for $P_1(t)$,
we thus obtain 
\begin{equation}
	P_1(s) = s^{-(\tau_s-1)} g(s\epsilon^{\nu_\parallel+d\nu_\perp-\beta}), 
		\qquad
		\tau_s = 1 + \frac{\beta_1}{\nu_\parallel+d\nu_\perp-\beta} .
					\label{eq:tau_1}
\end{equation}
This yields $\tau_s=1.287$ in nice agreement with our numerical result
in $1+1$ dimension. Compared to the value
$1+\beta/(\nu_\parallel+\nu_\perp-\beta)=1.108$ for DP with no boundary 
we find as expected that the boundary suppresses the cluster sizes.
{}From Eq.\ (\ref{eq:tau_1}) we find the average cluster size
\begin{equation}
	\left< s \right> \sim \epsilon^{-\gamma_1}, \qquad
		\gamma_1 = \nu_\parallel + d \nu_\perp - \beta - \beta_1 ,
					\label{eq:hyper}
\end{equation}
which replaces the usual hyperscaling relation 
when there is no wall present.

In standard DP the cluster mass and the order parameter scale with the same
exponent and there are three independent exponents.
In DP with an absorbing boundary, we find that the mass and the
order parameter scale with different exponents. One of these exponents
($\beta_1$) can however be related to the other exponents 
by assuming that $\tau_1=1$ and we are left
with the same three independent exponents as for standard DP in
$1+1$ dimensions.

\section{Boundary activity and backbone}

Now consider the activity on the absorbing boundary.
Direct measurement (by box counting) of 
the fractal dimension of living sites
along the time direction yields a dimension $D_{1}=0.578 \pm 0.005$.
This is an accordance with the result in figure~\ref{fig:boundary}
of an independent measurement
of the distribution of time intervals in which the boundary is dead, 
i.e., the distribution of first return $p_{\rm{first}}(t)$ of activity.
{}From figure~\ref{fig:boundary} we obtain 
$p_{\rm{first}}(t) \sim 1/t^{\tau_{\rm{first}}}$ with
$\tau_{\rm{first}}= 1.577 \pm 0.005$,
confirming the scaling formula $\tau_{\rm{first}}=1+D_1$.
We also obtain that the distribution of all return
$p_{\rm{all}}(t) \sim 1/t^{\tau_{\rm{all}}}$ fulfills
$\tau_{\rm{all}}=1-D_1$.
Without an absorbing boundary, the dimension of the activity along an 
arbitrarily longitudinal cut of the DP cluster would be 
$D_{\parallel}=1-\beta/\nu_{\parallel}=0.841$.
Thus, the activity on a cut
is much less when DP can evolve only on one side of this cut than
when the activity can move freely back and forth between both sides.
The connection between DP with a wall and the field-theoretic
formulation \cite{janssen-etal}
has been worked out in Ref.\ \cite{us}.
Therein it is shown that $\beta_1$ determines the scaling
of the density $\epsilon^{\beta_1}$ on the wall.
Thus, with the boundary we obtain the relation
$D_1 = 1-\beta_1/\nu_\parallel=1/\nu_\parallel$
in nice agreement with our numerical results.

The backbone is obtained from the infinite cluster by
removing all dangling ends. Thus the backbone consists of precisely 
those bonds which would be occupied by both the time-directed DP
process and its reversed time-directed process. It then follows that
the backbone density $\epsilon^{\beta^{BB}}$ is described by
the exponent $\beta^{BB} = 2\beta$.
We have numerically confirmed that this backbone behavior is still valid
for DP with a wall.  Besides, we have measured
the backbone dimension on the wall with the result
$D_{1}^{BB} = 1 - \beta_{1}^{BB}/\nu_\parallel = 0.16 \pm 0.01$.
This yields
$\beta_{1}^{BB} = 1.46$ in nice agreement with $\beta_{1}^{BB}=2\beta_1$,
cf.\ the result for the bulk.

On the backbone one can identify the so-called red bonds
\cite{stanley}
which, if one is cut, divide the cluster into two parts.
A renormalization group argument \cite{cognilio}
(see also \cite{huber}) yields for the number of red bonds up to scale
$t$ the scaling
\begin{equation}
	N_R(t) \sim t^{1/\nu_\parallel}.
		\label{eq:red-bonds}
\end{equation}
We have measured the scaling of red bonds for DP with a wall
and obtained results in complete agreement with Eq.~(\ref{eq:red-bonds}).
In addition we have measured the scaling
of the red bonds along a longitudinal cut for DP with no wall and
obtained the result $N_{R}^{cut}(t) \sim t^{-0.04 \pm 0.02}$.
This is in accordance with the expected behavior
$N_{R}^{cut}(t) \sim t^{1/\nu_\parallel - \chi} \sim t^{-0.056}$,
where the extra factor originates from the scaling of the width
$\xi_\perp$ of the cluster.
Finally, we have measured the scaling of red bonds on the wall for
DP with the wall and found the behavior
$N_{R,1}(t) \sim t^{-0.60 \pm 0.1}$.

\section{Results in $2+1$ dimensions and concluding remarks}

The above procedure can be applied to higher dimensions.
For $2+1$ dimensional bond directed percolation on a bcc lattice 
we have $p_c = 0.287338$,
$\beta=0.584$, $\nu_{\parallel}=1.295$, $\nu_{\perp}=0.734$, $\delta=0.451$,
and $\chi=\nu_{\parallel}/\nu_{\perp}= 0.567 $ \cite{2d,grass-zhang}.
We have measured the lifetime distribution, cluster distribution, and
average lifetime and cluster distributions. Again, we conclude that
$\nu_\parallel$ and $\nu_\perp$ are unchanged. Furthermore, $\beta$
is needed in order to describe the scaling of the mass whereas
a new exponent, $\beta_1$, is needed to describe the lifetime
distribution: we find the values $\beta_1 = 1.05 \pm 0.05$,
and $\tau_1 = 0.26 \pm 0.02$ in accordance with
$\tau_1 = \nu_\parallel - \beta_1$, cf.\ Eq.~(\ref{eq:P(t)}).
In contrast to the $1+1$ dimensional case, in $2+1$ dimensions
$\beta_1$ does not seem to be (simply) related to the other exponents
and we conclude that there are four independent exponents.
For $2+1$ dimensional systems with an edge (which introduces
yet another exponent) see Ref.\ \cite{us}.

One may speculate what the effect of the boundary 
will be if the cluster is initiated
some fixed distance $x$ away from the boundary.
In that case, scaling will be dominated by the DP behavior
(\ref{eq:P(t)})
until the meandering of the cluster allows it to reach the absorbing boundary,
i.e., until a crossover time
$t_\times \propto x^{\nu_{\parallel}/\nu_{\perp}}$. 
After this crossover time, the distribution of cluster survival times will
become steeper, and be determined by~(\ref{eq:P_1(t)}).
Thus the presence of a boundary will cause two scaling
regimes for the cluster size.
One can also consider the case where the DP process is initiated
in between two absorbing boundaries.
In such a case the DP process cannot evolve infinitely. After some time,
of the order of $L^{\nu_{\parallel}/\nu_{\perp}}$, where $L$ is the distance
between the boundaries, the otherwise critical DP process dies out.


In summary, we have investigated the effects of an absorbing boundary on DP
and discussed the novel scaling behavior originating from such a boundary.
The scaling can be understood in terms of the three exponents for DP
plus one new exponent. Our results are in agreement with recent series
expansion estimates. We also study the scaling at the boundary.
We stress that boundary effects are not limited to the case
where the boundary is absorbing---it could, e.g., also be reflecting.
The effect that reduces
the lifetimes of the clusters and the fractal dimension of activity
on the boundary is solely due to the fact that
living branches cannot meander back and forth over the boundary.
Finally, the exponent which determines the lifetime distribution in $1+1$
dimensions is $1/\nu_\parallel$, which describes the scaling
of the red bonds.
In the context of self-organized interface growth determined
by DP paths there exists a relation which involves $1/\nu_\parallel$
and dictates that the corresponding average `lifetime' should
scale as (\ref{eq:<t>}) with an exponent equal to unity
\cite{soc2}.
It would be interesting to explore these similarities and to possibly
relate the lifetime distribution to the scaling of the red bonds.

\section*{Acknowledgements}

We acknowledge discussions with Per Fr\"ojdh and Martin Howard;
the latter also for bringing Ref.\ \cite{janssen-etal} to our attention.
K.~B.~L.
acknowledges the support from the Danish Natural Science Research Council
and the Carlsberg Foundation.
M.~M. acknowledges the support from the Slovak Academy of Sciences 
(2/2025/97) and the PECO network program (CIPD-CT94-0011).

\end{document}